# Federating OMNeT++ Simulations with Testbed Environments

Asanga Udugama*, Koojana Kuladinithi†, Anna Förster*, Carmelita Görg*

*Sustainable Communication Networks Group, University of Bremen, Germany

Email: {adu|anna.foerster|cg}@comnets.uni-bremen.de

†Institute of Communication Networks, Hamburg University of Technology, Germany

Email: {koojana.kuladinithi}@tuhh.de

*Abstract*—We are in the process of developing a system architecture for opportunistic and information centric communications. This architecture (called Keetchi), meant for the Internet of Things (IoT) is focussed on enabling applications to perform distributed and decentralised communications among smart devices. To realise and evaluate this architecture, we follow a 3-step approach. Our first approach of evaluation is the development of a testbed with smart devices (mainly smart phones and tablets) deployed with this architecture including the applications. The second step is where the architecture is evaluated in large scale scenarios with the OMNeT++ simulation environment. The third step is where the OMNeT++ simulation environment is fed with traces of data collected from experiments done using the testbed. In realising these environments, we develop the functionality of this architecture as a common code base that is able to operate in the OMNeT++ environment as well as in the smart devices of the testbed (e.g., Android, iOS, Contiki, etc.). This paper presents the details of the "Write once, compile anywhere" (WOCA) code base architecture of Keetchi.

## I. INTRODUCTION

The use of networks is moving in the direction of being information centric. International Data Corporation (IDC) predicts, that by 2020, the digital universe will have close to 45 ZB of data [1]. In the same report of current and future trends, IDC predicts further that by 2020, the total number of connectable devices in the Internet of Things (IoT) will be above 200 billion. To address these trends and the different application areas, new communication architectures and protocols have to be designed. One such architecture that we are building is Keetchi. It addresses the opportunistic communications paradigm with the focus on distributed, decentralised and reinforcement learning based [2] [3] communications.

Keetchi is meant to operate in a number of different environments focusing on large scale networks. Firstly, in infrastructure-less environments, nodes deployed with the Keetchi architecture will perform direct communications with other nodes. Secondly, when infrastructure is present, nodes may use the infrastructure to reach other nodes. Thirdly, Keetchi based nodes may also operate in hybrid environments where infrastructure-less environments interact with infrastructure based environments. When developing this architecture and its functionality, our intensions go beyond designing and evaluation of efficient mechanisms. On the one hand, we wish to develop code that is platform independent and is able to be executed not only in OMNeT++, but also on real IoT based devices. Currently, we are building a testbed consisting of hundreds of IoT based devices installed with Android, iOS, Contiki and Embedded Linux operating systems. On the other hand, we are also interested in large scale evaluations, which can only be realised with simulations. At the same time, simulations need to be sophisticated with traces from real applications and environments to make the results more realistic and representable.

Therefore, in this paper we present our on-going work on developing a platform independent code base of Keetchi that enables distributed, decentralised and reinforcement learning based communications in OMNeT++ as well as in the smart devices deployed in a testbed. As the focus of this paper is the code base, we do not describe the Keetchi architecture in detail. But, a brief description of the salient aspects of the Keetchi architecture is provided in the next section to assist the understanding of the subsequent sections.

## II. SYSTEM ARCHITECTURE OF KEETCHI

At the heart of Keetchi is the emphasis on exploiting the benefits of information centricity of today's communications. It has the following main features:

- **Information Centric Communications** - Consideration of named data as in Content Centric Networking (CCN)[4] and placing less emphasis on naming hosts (current networks),
- **Distributed Control** - Every device is in charge of its own operations without any dependance on centralised controls,
- **Distributed Caching** - Devices capable of storing data select *when* and *what* to store based on local decisions,
- **Reinforcement Leaning Based Data Handling** - Feedback for exchanged data decides what data is considered for propagation in the network and what data is stored in caches,
- **Opportunistic Data Propagation** - Constant monitoring of the neighbourhood (i.e., to build a connectivity map) is performed to identify opportunities for communications and infrastructure-less communications.

The Keetchi architecture which is deployed in mobile devices (i.e., Keetchi nodes) and used by the applications in those mobile devices, consists of a number of modules that interact to store and propagate data generated by the applications.

Figure 1 shows an example of the message propagations in the Keetchi architecture in the time domain. *Node A* propagates data for which feedback messages are received (*Node B* and *Node C*). Once *Node B* meets *Node E*, it can re-propagate the data as *Node E* had previously indicated a preference for such data. The scenario shown in Figure 1 is for an infrastructure-less deployment which could employ bearer technologies such as Wifi Direct or Bluetooth Low Energy (BLE) to communicate in a peer-to-peer communication model. A deployment in an infrastructure based environment of Keetchi may use WiFi or a 3rd Generation Partnership Program (3GPP) technology.

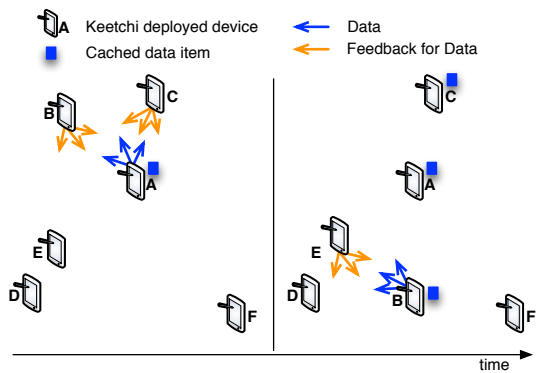

Fig. 1. Operations in a Keetchi based network

Every Keetchi node has a 3-layer protocol stack. These layers perform the following activities.

- **Application Layer** - The application layer provides the support required to host applications that are based on decentralised and distributed direct communications.
- **Keetchi Layer** - The Keetchi layer handles the functions of maintaining the different data dissemination strategies together with the communication models employed and data caching based on the features detailed above.
- **Link Layer** - The Keetchi architecture leverages link technologies such as Wi-Fi Direct or BLE to perform direct communications.

Communication in the Keetchi architecture is performed using 2 message types. They are:

- **Data Message** - The Data message[1] carries the named data generated by applications executed in nodes using the Keetchi architecture. These messages are propagated through the network opportunistically and controlled by the reinforcement learning caching and forwarding model.
- **Feedback Message** - The Feedback message[1] carries information that evaluates previously received named data or as a means of soliciting additional data. Its serves as a reward to the reinforcement learning model.

The work presented in [5] describes a model and the evaluation of a peer-to-peer content-centric model for multi-group communications with WiFi Direct. [6] presents a study on feasibility, advantages and challenges of an IoT based on Named Data Networking (NDN). The Keetchi architecture includes some of the aspects considered in [5] and [6] (i.e., information centric communications, IoT, WiFi Direct, etc.). But the differentiating aspect of Keetchi with the aforementioned work is its additional focus on opportunistic communications and reinforcement learning based data handling which are vital for communications in the IoT.

## III. CODE BASE ARCHITECTURE

The Keetchi architecture is envisioned to be deployed in a number of different computing platforms. They range from ubiquitous devices such as smart phones to custom built computing devices for rugged environments. Additionally, the Keetchi architecture is envisaged to be evaluated for its performance in large scale simulations in OMNeT++.

To cater to these 2 requirements, i.e., deployment in testbed environments and evaluation in OMNeT++, the functionality related to performing the operations unique to the Keetchi layer are developed as a common codebase. Figure 2 shows the architecture of the codebase with some example platforms on which it is expected to be executed.

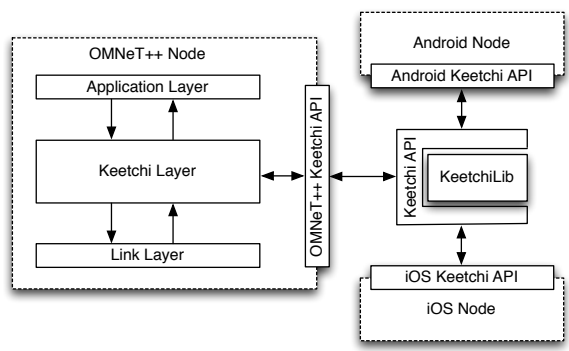

Fig. 2. Codebase Architecture

The functionality of the Keetchi layer operations are placed in a library, *KeetchiLib*. Each computing platform has an API that calls the functionality of Keetchi operations. The KeetchiLib is developed using C++ and the APIs handle the translations between C++ and the native languages used in those computing platforms. For example, in the Android platform, the API converts between Java and the C++ code using the Java Native Interface (JNI) functionality. Depending on the architecture of how applications are enabled in a specific computing platform, the KeetchiLib is used as a dynamic or a static library.

With OMNeT++, the API is a simple pass through API and the KeetchiLib could be configured to be used either as a dynamic or a static linked library.

The KeetchiLib consists of a number of classes that are instantiated and released during operations. The main class, KeetchiMain, is instantiated at and held during the lifetime of each single Keetchi node. The Keetchi Layer of each of the computing platforms must instantiate the KeetchiMain class to access all Keetchi functionality. This functionality consists of code to handle data store operations and a set of algorithms including other functions (described later). The data stores are implemented as hash maps to store the data that traverse the Keetchi node and a map of neighbours (all nodes met

---

[1]The term *Data* message has the same semantic meaning as in CCN with the difference being (due to distributed infrastructure-less communications) that Data messages are sent out opportunistically without receiving a CCN *Interest* message for this data first. Due to these differences the second type of message used in Keetchi is called the *Feedback* message which reverses the way interests for data are announced: it evaluates already received data in order to either reinforce its dissemination or inhibit it. This forms the basis of the Keetchi reinforcement learning algorithm.

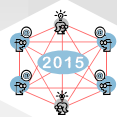

```
/**
 * Process an incoming Feedback message from either the application
 * layer or the link layer.
 *
 * @param fromWhere    The message source, application layer or link
 *                     layer.
 * @param feedbackMsg  The Feedback message with its contents as a
 *                     KLFeedbackMsg instance.
 * @param currentTime  The current clock time.
 * @return             A KLAction instance with all actions required
 *                     to be taken by the Keetchi layer.
 */
KLAction* processFeedbackMsg(int fromWhere, KLFeedbackMsg *feedbackMsg,
                             double currentTime);
```

Fig. 3.   An example method of the KeetchiMain class

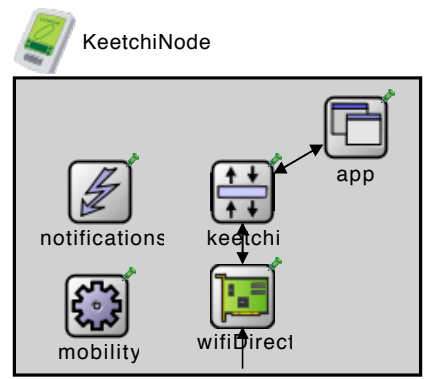

Fig. 4.   OMNeT++ Keetchi Node Architecture

up to now). The algorithms in the KeetchiLib are for cache management, for execution of the dissemination strategies and for activating the different communication models.

Figure 3 shows a method definition of one of the functions of the KeetchiMain class. This method processes an incoming Feedback message from either the application layer or the link layer (*fromWhere*). Once processed, the method returns a KLAction type instance with instructions on what actions must be taken by the Keetchi layer and the information required for this. These actions may range from taking no action to sending multiple messages to the application and link layer.

An important aspect in the KeetchiLib is the use of the current clock time. The clock time is required in multiple activities of Keetchi such as for time stamping messages. Since simulators operate on the simulation time, the system clock is never used in KeetchiLib. Instead, the current time is passed in all methods (where it is required) as a parameter.

The functionality exposed in the KeetchiLib for OMNeT++ fall mainly into 4 categories. These are as follows:

- **Incoming Message Processing** - As described above, operations at Keetchi nodes result in the arrival of Feedback and Data messages at other Keetchi nodes. These messages trigger the Keetchi layer to perform other actions.
- **Opportunistic Message Generations** - In addition to the above messages (Feedback and Data messages), another trigger for activity at the Keetchi layer is the receipt of neighbourhood node information. These will trigger the generation of Data messages.
- **Status Information Servicing** - KeetchiLib creates and maintains a number of data stores in its operations. This set of functions provide access to the information held by these data stores.
- **Wiring for Statistics** - The functions related to dumping statistical information is coupled with the functions of the previous function categories. The return action information of the previous functions also bring back statistical information that could be used to generate statistics in OMNeT++.

## IV. KEETCHI OMNET++ MODEL

A Keetchi node in OMNeT++ implements the 3-layer protocol stack (shown in Figure 2) together with the use of mobility models, traffic models (application events) and wireless propagation models (physical layer) available in OMNeT++. Figure 4 shows the implementation of the protocol stack and the connections between the protocol layers of a Keetchi node in OMNeT++. There are a number of applications considered for the Keetchi environment. They range from emergency messaging applications to social networking. There are 3 types of applications in general. *Traffic Generators* produce data that expect feedback to make decisions on the type of data to generate and the recipients to send to, subsequently. *Traffic Consumers* are the users of data who generate feedback based on preferences. The third type of applications are the hybrid applications that perform both of the above (send and receive).

When considering sources of traffic (application events), we consider applications that generate traffic based on the synthetic models available in OMNeT++ or, based on the traces obtained from our IoT testbed.

The Keetchi layer of a node is responsible for providing the Keetchi functionality through the KeetchiLib. Using the *OMNeT++ Keetchi API*, the Keetchi layer creates an instance of a KeetchiMain class during the intitialisations (*initialize()*) using a multi-stage initialisation process. The Keetchi layer will receive Feedback and Data messages from the applications and the Link layer which are passed on to the OMNeT++ Keetchi API for processing. The C++ class in OMNeT++ for the Keetchi layer (*Keetchi.cc*) converts the properties in the messages into the format of the message required by the KeetchiLib. Once processed, KeetchiLib will return the actions to be taken by the OMNeT++ Keetchi layer. These actions may result in generating multiple messages to be sent to either the applications, the Link layer or both.

Another type of information passed on to the KeetchiLib is the information on nodes currently reachable in the neighbourhood. This information, which is received from the OMNeT++ Link layer models is used by the KeetchiLib to build a map of the current nodes in the vicinity and to generate Data messages which have been previously indicated as being of interest (opportunistic communications).

The Link layer in the Keetchi architecture is realised using a Wi-Fi Direct bearer. The WiFi Adhoc implementation of the INET framework of OMNeT++ is being extended to include the functionality related to Wi-Fi Direct operations. In addition to providing functionality to establish connections with multiple other Keetchi nodes, the Wi-Fi Direct extensions also include the provisioning of information of the nodes currently in the neighbourhood. As described above, the Keetchi layer is overlaid over a Wi-Fi Direct Link layer. Therefore, the Keetchi layer uses the Media Access Control (MAC) addresses of nodes to identify a node uniquely.

## V. A SIMULATED SCENARIO AND EVALUATION METRICS

The Keetchi architecture is envisioned to be deployed in a number of application areas. These areas relate to emergency services, social networking, smart building controls, etc. One such application is an exchange service for "hand-me-down" goods. This application, called *UniRecycler* allows users to inform about goods that are available to give away and for users who are interested in these goods to inform back of their interest and feedback.

The UniRecycler application uses the peer-to-peer, decentralised and distributed communication model of the Keetchi architecture. Additionally, it uses data caching and opportunistic communications of the architecture to perform store-and-forward communications. When evaluating such an application in OMNeT++, we consider large scale network scenarios with many individual devices. The nodes in this scenario would connect to other nodes in their vicinity, perform communications, and disconnect again due to mobility or other issues. The nodes that move away may act as carriers of data collected on behalf of other nodes, to be transferred if they connect later.

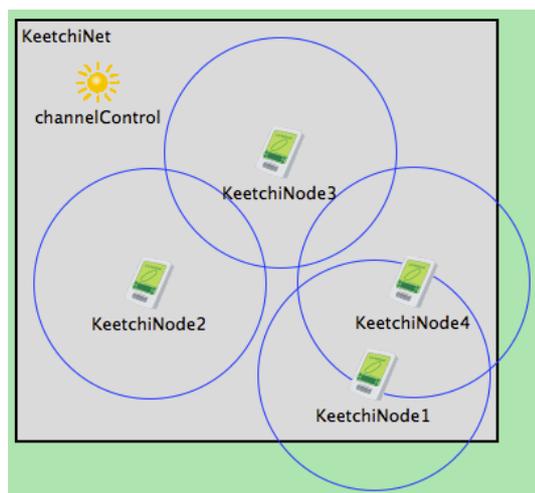

Fig. 5. An OMNeT++ Scenario with Keetchi Nodes

Figure 5 shows the operation of a small scenario with multiple nodes deployed with the Keetchi architecture. The user of *KeetchiNode1* may inform of a give-away through a Data message which gets propagated to the *KeetchiNode2* over the Wi-Fi Direct based connections of *KeetchiNode4* and *KeetchiNode3*. When the Data message is received, some of these nodes (e.g., *KeetchiNode3*) may send a Feedback message indicating its interest in such data. Further, if *KeetchiNode2* decides to disconnect and connect to other clusters, it may decide to resend the original Data message from its cache based on previously received feedback for such data.

When considering large scale scenarios, evaluations in testbed environments are insufficient due to the limitations of the scale related issues of testbeds (e.g., number of devices, connection ranges, etc). Moreover, even in small scale networks that could be deployed with testbeds, limitations of the used devices may hamper proper evaluations (e.g., availability of only one type of bearer technology).

On the other hand, OMNeT++ based simulated scenarios provide the flexibility to create networks with extremely large node numbers spanning large geographical areas and the ability to have full-featured devices for evaluations. Evaluating the Keetchi architecture in such scenarios is required to understand the behaviour of mechanisms used in all parts of the Keetchi architecture. Therefore, we consider a number of evaluation metrics that focus on the following areas:

**Feedback System**: Feedback messages carry a numeric weight that is computed at each node. This metric should consider the popularity, freshness, location and the usefulness of data. Algorithms such as Q-learning are employed by the Keetchi architecture to compute these weights. Large scale evaluations provide the possibility of understanding the influences on the weights depending on the node density and mobility.

**Opportunistic Data Propagation Strategies**: One of the novelties of the Keetchi architecture is its self adaptiveness employed in propagating data depending on the available communication opportunities and feedback from neighbours. The ability of OMNeT++ based scenarios to consider full-featured devices (e.g., devices with Wi-Fi Direct and BLE together) in simulations has the possibility of evaluating the performance of opportunistic communication over multiple bearers simultaneously.

**Caching**: Distributed caching, cache policies, cache sizes and the activities of nodes influence what data is ultimately held in caches. The OMNeT++ based scenarios considering scale networks and full-featured devices (e.g., larger caching capabilities) with varying traffic patters help in understanding the issues of scalability in Keetchi architecture based networks.

**Mobility Patterns**: There are many mobility patterns to consider when evaluating the Keetchi architecture. These patterns depend not only on the application type but also on many other factors (e.g., contact time, data prioritisation, etc.) Due to the flexibility of configuration, the influence on performance of these mobility patterns are best evaluated in a simulated environment compared to testbed environments.

## VI. TECHNICAL CHALLENGES

The Keetchi layer which implements the Keetchi architecture is impervious of the underlying networking bearer technologies. The underlying link layer technology used is, in general, required to perform hop-by-hop, peer-to-peer communications and provide information of the neighborhood. In our work, we focus on WiFi Direct and BLE technologies. There are a number of challanges that have to be addressed to use these technologies in the Keetchi architecture. In the following is a brief description of these challenges.

**WiFi Direct** - WiFi Direct is a standard of the WiFi Alliance to perform peer-to-peer (or multi-peer) communications. It is an extension to the functionality of WiFi. WiFi Direct requires the establishment of a trusted connection between peers before any communications could occur. Current implementations of WiFi Direct (mainly in the Android platform) require re-exchanging of trust credentials through user interactions when re-establishing communication sessions. When exchanging Keetchi messages (using WiFi Direct), we consider the re-exchanging of trust credentials to be a drawback .

**Bluetooth Low Energy** - Bluetooth Low Energy (BLE) provides a very energy efficient means of performing peer-to-peer communications. When compared to the operations of Bluetooth, the available bandwidth of BLE is lower. BLE is especially meant for small message exchanges. But, when considering the communication patterns of Keetchi, there is a requirement to transmit messages with large payloads (e.g., images). Therefore, the link layer used by the Keetchi architecture must be able to select the bearer based on the type of message to send (i.e., cross-layer control).

## VII.  Conclusion

Services envisaged for the IoT require the development of new communication architectures. Keetchi is a communication architecture meant for IoT based ubiquitous devices to host some of these services. The Keetchi architecture, which exploits the benefits of opportunistic communications and the information centricity of today's network use, provides the means for applications to perform peer-to-peer, distributed and decentralised communications. When developing and evaluating such an architecture, our focus has been to develop a code base of the functionality that is able to be used not only on the implementations for the different operating systems but also in OMNeT++. We are in the process of building such a code base and this paper presents this on-going work in the context of OMNeT++. By presenting this work, firstly, we wish to share our experiences and secondly, to discuss with the community of OMNeT++ users and developers about the features that we intend to build (e.g., mobility models for opportunistic communications, WiFi Direct and BLE) which are mutually beneficial for our work as well as for the community.